\documentclass[sn-mathphys,Numbered]{sn-jnl}

\usepackage{graphicx}%
\usepackage{multirow}%
\usepackage{amsmath,amssymb,amsfonts}%
\usepackage{amsthm}%
\usepackage{mathrsfs}%
\usepackage[title]{appendix}%
\usepackage{xcolor}%
\usepackage{textcomp}%
\usepackage{manyfoot}%
\usepackage{booktabs}%
\usepackage{algorithm}%
\usepackage{algorithmicx}%
\usepackage{algpseudocode}%
\usepackage{listings}%
\usepackage{aas_macros}
\usepackage[]{hyperref}
\usepackage[normalem]{ulem}

\def\aap{\textit{Astronomy and Astrophysics}}

\def\be{\begin{equation}}
\def\ee{\end{equation}}
\def\bea{\begin{eqnarray}}
\def\eea{\end{eqnarray}}

\newcommand\Ol{\mathcal O}

\def\L{\mathcal{L}}
\def\Lm{\mathcal{L}_m}

\newcommand\D{\mathcal D}

\begin{document}

\title[Slowly rotating and charged Black-holes in Entangled Relativity]{Slowly rotating and charged Black-holes in Entangled Relativity}


\author[1,3]{\fnm{Maxime} \sur{Wavasseur}}\email{m.wavasseur@protonmail.com}

\author[2]{\fnm{Théo} \sur{Abrial}}\email{theo.abrial@ens-paris-saclay.fr}

\author*[3,4]{\fnm{Olivier} \sur{Minazzoli}}\email{ominazzoli@gmail.com}

\affil[1]{\orgdiv{Departament de Física Quántica i Astrofísica (FQA)}, \orgname{Universitat de Barcelona (UB)}, \orgaddress{\street{Carrer de Martí i Franquès, 1}, \city{Barcelona}, \postcode{08028}, \country{Spain}}}

\affil[2]{\orgdiv{Département de Physique}, \orgname{ENS Paris-Saclay}, \orgaddress{\street{4 Avenue des Sciences}, \city{Gif-sur-Yvette}, \postcode{91190}, \country{France}}}

\affil[3]{Universit\'e C\^ote d'Azur, Observatoire de la C\^ote d'Azur, CNRS, Artemis, Boulevard de l'Observatoire, 06304, Nice, France}
\affil[4]{Bureau des Affaires Spatiales, 2 rue du Gabian, 98000,  Monaco}


\abstract{
Entangled Relativity is a non-linear reformulation of Einstein's General Theory of Relativity (General Relativity) that offers a more parsimonious formulation. This non-linear approach notably requires the simultaneous definition of matter fields, thus aligning more closely with Einstein's \textit{principle of relativity of inertia} than General Relativity does. Solutions for spherically charged black holes have already been identified. After exploring further some of the properties of these solutions, we present new solutions for the field equations pertaining to slowly rotating charged black holes.
}




\maketitle

\tableofcontents 
\section{Introduction}\label{sec:intro}

Entangled Relativity is a general theory of relativity that intertwines curvature (gravity) and matter fields in the definition of the theory through a non-linear coupling \cite{ludwig:2015pl,arruga:2021pr}. Therefore, it better aligns with Einstein's original opinion that a satisfying relativistic theory should prevent the existence of vacuum solutions, which he named the \textit{principle of relativity of inertia},\footnote{Also named \textit{Mach's principle} by Einstein in \cite{einstein:1918an}. He explains this choice as follows: ``I have chosen the term “Mach’s principle” because this principle is a generalization of Mach’s claim that inertia has to be reduced upon interaction of the bodies''.} and which notably implies, in Einstein's words, that ``there can be no [spacetime metric] field without matter'' \cite{einstein:1918an}---see also \cite{einstein:1917co,einstein:1918sp,hoefer:1995cf,pais:1982bk}. Indeed, because of the non-linear coupling between matter and curvature in the formulation of the theory, it cannot even be defined without matter fields.\footnote{See \cite{minazzoli:2024pn} for a broader discussion on Mach's principle in both General Relativity and Entangled Relativity.} Also, because it requires less universal dimensionful constants than General Relativity in order to be defined, it is also more parsimonious than General Relativity---both at the classical and at the quantum levels \cite{minazzoli:2022ar,minazzoli:2023ar}. We will see this point with the formulation of the theory below.\\ 

Indeed, Entangled Relativity can be defined from the following path integral \cite{minazzoli:2022ar,minazzoli:2023ar}
\be
Z= \int [\D g]  \prod_i [\D f_i] \exp[i\Theta], \label{eq:Entangled RelativityPI}
\ee
with 
\be \label{eq:Entangled RelativityQP}
\Theta = -\frac{1}{2 \epsilon^2} \int d^4_g x \frac{\L^2_m(f,g)}{R(g)}, 
\ee
and $\int [\D]$ relates to the sum over all possible (non-redundant) field configurations, $R$ is the usual Ricci scalar that is constructed upon the metric tensor $g$, $ \mathrm{d}^4_g x := \sqrt{-|g|}  \mathrm{d}^4 x$ is the spacetime volume element, with $|g|$ the metric $g$ determinant, and $\L_m$ is the Lagrangian density of matter fields $f$. It is not necessarily the Standard Model of particles Lagrangian density at the fundamental level, and could be a completion of it instead. It also depends on the metric tensor, a priori through to the usual \textit{comma-goes-to-semicolon rule} \cite{MTW}---at least it has to be the case in the limit where Entangled Relativity reduces to General Relativity. Because the dimension of the integral is an energy squared, and in order to recover standard quantum field theory on ``\textit{flat spacetime}'' when gravity is neglected,\footnote{A ``\textit{flat spacetime}'' is only a useful approximation on scale for which gravity can be neglected. This approximation fails both at large scales and at the infinitesimal scale of quantum gravity. Classically, even for large $R$, there always exists a given scale at which the metric can be approximated to be flat. Nevertheless, according to present knowledge, $R\neq 0$ everywhere in our universe.} $\epsilon$ is the (reduced) Planck energy \cite{minazzoli:2022ar,minazzoli:2023ar}. One can check that the Planck energy does not influence the classical phenomenology of the theory, which corresponds to stationary quantum phase $\delta \Theta = 0$.\\


An important characteristic of Entangled Relativity is that, despite its non-linear formulation, it has a built-in decoupling---originally found in scalar-tensor theories and named \textit{intrinsic decoupling} \cite{minazzoli:2013pr}---such that its phenomenology  is very close to that of General Relativity whenever matter fields are such that $\Lm \approx T$ on-shell. Since this is the case for typical matter fields present in our universe, it means that Entangled Relativity is not expected to deviate much from General Relativity for the predictions of several phenomena. In the Solar System for instance, Entangled Relativity is indistinguishable from General Relativity with current observational accuracy \cite{ludwig:2015pl}, or more generally at larger scales, due to the main contribution of non, or weakly, interacting particles that are supposed to constitute cold dark-matter.\\

Various observational campaigns, involving either the detection of gravitational waves with the LIGO, Virgo and later KAGRA collaborations \cite{GW150914}, or the reconstruction of the  image of a supermassive black-hole's shadow through interferometry \cite{EHT:2019aj}, have yielded data that are consistent with vacuum solutions of black hole in General Relativity, such as the Schwarzschild and Kerr metrics. Therefore, it is crucial that Entangled Relativity reproduces, at least up to non-observable differences, the predictions of Einstein's theory for astrophysical black-holes.\\

Minazzoli \& Santos \cite{minazzoli:2021ej} demonstrated in the case of a non-rotating charged black hole that the external Schwarzschild metric can be a good approximation of the exact solutions of black holes in Entangled Relativity in astrophysical conditions. They used the electromagnetic field as the sole source of matter, being the simplest choice for matter fields with infinite range. The natural extension of this study is to look at the case of rotating charged black holes, which is what the present paper will mainly be about. 
\section{Field equations}\label{sec:FE}

Classical physics corresponds to the paths in the path integral that lead to a stationary quantum phase $\delta \Theta = 0$. The reason being that for classical, or ``\textit{macroscopic}'', phenomena, destructive interference cancel any contribution from other paths to the path integral, whereas constructive interference are maximal for paths that lead to a stationary phase. This is usually referred to as the \textit{Principle of Least Action}, because if $\hbar$ is a fundamental constant, then $\Theta = S / \hbar$ and $\delta \Theta = 0 \Leftrightarrow \delta S = 0$, where $S$ has the dimension of an action. However, $\hbar$ is not a constant in Entangled Relativity\footnote{Trivially, $\hbar$ does not appear in the formulation of the theory in Eq. (\ref{eq:Entangled RelativityPI}).} \cite{minazzoli:2022ar,minazzoli:2023ar} and the quantum phase in Eq. (\ref{eq:Entangled RelativityPI}) is defined from an integral $E2 :=-1/2 \int d^4_g x \Lm^2/R$ with the dimension of an energy squared, such that $\Theta = E2/\epsilon^2$ instead, where $\epsilon^2$ is a universal constant energy squared. 
\subsection{Original form}

Therefore, the field equations in Entangled Relativity follow from the extremization of the quantum phase $\Theta$ in Eq. (\ref{eq:Entangled RelativityPI}). In this expression, we clearly see why it no longer makes sense to define any dynamics in absolute vacuum: the theory cannot even be defined in that case. This is illustrated in the metric field equation as well, which reads \cite{ludwig:2015pl}:

\be
R_{\mu\nu}-\frac{1}{2}g_{\mu\nu}R=-\frac{R}{\mathcal{L}_{m}}T_{\mu\nu}+\frac{R^{2}}{\mathcal{L}_{m}^{2}}\square\left(\nabla_{\mu}\nabla_{\nu} - g_{\mu\nu}\square\right)  \frac{\mathcal{L}_{m}^{2}}{R^{2}}, \label{eq:fRmetricfield}
\ee
with
\be
T_{\mu \nu} \equiv-\frac{2}{\sqrt{-g}} \frac{\delta\left(\sqrt{-g} \mathcal{L}_{m}\right)}{\delta g^{\mu \nu}}.
\ee
Let us note that the stress-energy tensor is not conserved in general (see \cite{minazzoli:2021ej}), as one has
\be
\nabla_{\sigma}\left(\frac{\mathcal{L}_{m}}{R} T^{\alpha \sigma}\right)=\mathcal{L}_{m} \nabla^{\alpha}\left(\frac{\mathcal{L}_{m}}{R}\right). \label{eq:noconsfR}
\ee
The matter field equation, for any tensorial matter field $\chi$, gets modified due to the non-linear coupling between matter and curvature as follows
\be
\frac{\partial \mathcal{L}_{m}}{\partial \chi}-\frac{1}{\sqrt{-|g|}} \partial_{\sigma}\left(\frac{\partial \sqrt{-|g|} \mathcal{L}_{m}}{\partial\left(\partial_{\sigma} \chi\right)}\right)=\frac{\partial \mathcal{L}_{m}}{\partial\left(\partial_{\sigma} \chi\right)} \frac{R}{\mathcal{L}_{m}} \partial_{\sigma}\left(\frac{\mathcal{L}_{m}}{R}\right). \label{eq:Entangled Relativitymatter}
\ee
In particular, it corresponds to a special case of $f(R,\Lm)$ theories \cite{harko:2010ep}. As usual in these theories, the differential equation on the extra-degree of freedom is given by the trace of the metric field Eq. (\ref{eq:fRmetricfield}), it reads
\be
3\square \frac{\mathcal{L}_{m}^{2}}{R^{2}}=-\frac{\Lm}{R}\left(T-\Lm\right). \label{eq:fRmetricfieldt}
\ee
This means that while $R/T$ is a constant in General Relativity, $R/\Lm$ is a field in Entangled Relativity.\\


It is worth noting that if $\Lm = T$, then the equation for $R/\Lm$ remains unsourced, allowing $R/\Lm = R/T$ to be constant, akin to what happens in General Relativity, where $R/T = - 8\pi G/c^4$. The precise cancellation on the right-hand side of Eq. (\ref{eq:fRmetricfieldt}) when $\Lm=T$ has been termed an \textit{intrinsic decoupling} in \cite{minazzoli:2013pr}---in the context of scalar-tensor theories.

It is worth noting that Entangled Relativity shares many of its predictions with General Relativity, and usually only differs in high-energy-density regimes \cite{arruga:2021pr,arruga:2021ep}. Indeed, most of the material content of our universe at the present cosmological epoch is such that $\Lm \approx T$. This is the case for instance in the Solar System \cite{minazzoli:2013pr} or more generally at larger scales, due to the main contribution of non, or weakly, interacting particles that are supposed to constitute cold dark-matter.
Therefore, the solutions of General Relativity generally constitute very good approximations of those of Entangled Relativity (\cite{minazzoli:2021ej}). Indeed, the often extremely weak variation of the additional degree of freedom provided by the ratio between $\Lm$ and $R$ in the equation implies a phenomenology that closely mimics that of Einstein's theory \cite{minazzoli:2018pr,minazzoli:2022ar}.

\subsection{An equivalent form}

Quite intriguingly, these equations can also be derived from the following (pretty standard) Einstein-dilaton phase:
\be \label{eq:Entangled Relativity2QP}
\Theta_{Ed} = \frac{1}{\epsilon^2} \int d^4_g x \frac{1}{\kappa}\left(\frac{R(g)}{2 \kappa} + \L_m(f,g) \right),
\ee
provided that $\L_m \neq \emptyset$,\footnote{If $\L_m = \emptyset$ in Eq. (\ref{eq:Entangled Relativity2QP}), then it does not correspond to Entangled Relativity, as defined in Eq. (\ref{eq:Entangled RelativityQP}).} and where $\kappa$ is a scalar-field. Obviously, the value of $\epsilon$ does not impact the classical limit of the theory. The classical equivalence between the original \textit{$f(R,\L_m)$} theory in Eq. (\ref{eq:Entangled RelativityPI}) and the \textit{Einstein-dilaton} theory in Eq. (\ref{eq:Entangled Relativity2QP}), stems from a very well known fact: non-linear algebraic functions of the Ricci scalar in the action are equivalent to having an additional scalar degree-of-freedom with gravitational strength \cite{capozziello:2015sc,teyssandier:1983jm,jakubiec:1988pr}. As a consequence, it indicates that the theory defined in Eq. (\ref{eq:Entangled RelativityPI}) should be immune to the Ostrogradskian instability and to the non-well-posedness of the Cauchy problem despite not being of second order---just as $f(R)$ theories \cite{woodard:2007ln,teyssandier:1983jm,jakubiec:1988pr}. \\

Indeed, the Euler-Lagrange equation for the scalar-field implies
\be \label{eq:defkappa}
\kappa = -\frac{R}{\L_m},
\ee
instead of $\kappa = - R/T$ in General Relativity. Using Eq. (\ref{eq:defkappa}), the trace of the metric field equation (\ref{eq:fRmetricfieldt}) can be rewritten as follows
\be
3 \kappa^{2}\square \kappa^{-2}=\kappa \left(T-\mathcal{L}_{m}\right). \label{eq:sceq}
\ee

One may think that Eq. (\ref{eq:defkappa}) is singular for $\Lm \rightarrow 0$. However, this is not correct, as Eq. (\ref{eq:defkappa}) is not independent from the other field equations. In particular, the differential equation that commands the behavior of the ratio between $R$ and $\Lm$ is given by Eq. (\ref{eq:sceq})---which is a perfectly well-behaved equation. In some sense, it is like for the ratio between $R$ and $T$ in General Relativity that is well-behaved at $T \rightarrow 0$ because the equations of General Relativity is such that $R\propto T$.  In what follows, we will see how it works in particular when the charge of a black-hole goes to zero, and verify that $R/\Lm$ indeed remains finite at the limit $\Lm \rightarrow 0$.

Let us note that one can use an adimensional scalar-field $\phi$, or $\vartheta$, instead with
\be
\sqrt{\phi} = \frac{\bar \kappa}{\kappa},
\ee
or
\be
\vartheta= \frac{\bar \kappa}{\kappa},
\ee
 where $\bar \kappa$ is a dimensionful normalisation constant, which would depend on the specific background value or initial conditions that one would consider. With this parametrization, Eq. (\ref{eq:Entangled Relativity2QP}) would simply read as follows
\be \label{eq:Entangled Relativity2phi}
\Theta_{Ed} = \frac{1}{\bar \kappa \epsilon^2} \int d^4_g x \left(\frac{\phi R(g)}{2 \bar \kappa} + \sqrt{\phi} \L_m(f,g) \right),
\ee
or
\be \label{eq:Entangled Relativity2vartheta}
\Theta_{Ed} = \frac{1}{\bar \kappa \epsilon^2} \int d^4_g x \left(\frac{\vartheta^2 R(g)}{2 \bar \kappa} + \vartheta \L_m(f,g) \right).
\ee

Let us note that the formulation with $\vartheta$ is more general because it allows, a priori, for negative Newton's parameter, i.e. $\vartheta <0$ or $\kappa <0$. In what follows, we consider units that are such that $\bar \kappa = c = \mu_0= 1$, where $\mu_0$ is the magnetic permeability---such that one notably has
\be
\sqrt{\phi} = 1/ \kappa=\vartheta.
\ee

\subsection{Electromagnetism}
Here we consider the simplest case of matter field with infinite range, where the Lagrangian density consists solely of the electromagnetic field 
\be
\Lm = - \frac{1}{2} F_{\mu \nu} F^{\mu \nu},
\ee
with $F_{\mu \nu}=\partial_\mu A_\nu - \partial_\nu A_\mu$ and $A^\alpha$ the four-vector electromagnetic potential. The stress-energy tensor is
\be
T_{\mu \nu} = 2 \left(F_{\mu \sigma} F_\nu^{~\sigma} + g_{\mu \nu} \frac{\Lm}{2}\right).
\ee
The electromagnetic field equation on the other hand reads
\be
\nabla_\sigma \left(\frac{F^{\mu \sigma}}{\kappa} \right) = 0.\label{eq:Maxwell}
\ee
Let us note that, despite a different field equation, electromagnetic waves still follow null-geodesics of spacetime in the geometric optic approximation \cite{minazzoli:2013pr,minazzoli:2014pr}.

\section{Spherically (non rotating) charged BH}

As we mentioned in the introduction, the non-rotating case has already been addressed by \cite{minazzoli:2021ej}. They found for a spherically electrically charged black-hole, the metric field solution reads 

\be
ds^2 = -\lambda_0^2 dt^2 + \lambda_r^{-2} dr^2 + \rho^2 d\Omega^2,\label{eq:metric}
\ee
with
\bea
&&\lambda_0^2 = \left(1-\frac{r_+}{r} \right)\kappa^{15/2},\label{eq:NRBH_lambda0}\\
&&\lambda_r^2 = \left(1-\frac{r_+}{r} \right)\kappa^{7/2},\\
&&\rho^2 = r^2 \kappa^{3},\label{eq:rho}
\eea
while the scalar-field solution reads
\be \label{eq:kappa}
\kappa = \frac{1}{\sqrt{\phi}} = \frac{1}{\vartheta} = \left(1-\frac{r_-}{r} \right)^{2/13}
\ee
and the electric field solution is
\be \label{eq:Asol}
A=-\frac{Q}{r} dt
\ee
The mass $M$ and charge $Q$ of the black-hole can be recovered according to
\bea \label{eq:masse-charge}
&&2M = r_+ + \frac{11}{13} r_- \label{eq:Mass}\\
&&Q^2 = \frac{12}{13} r_+ r_- \label{eq: Charge}
\eea

The parameters $r_{+}$ and $r_{-}$ define an event horizon and a curvature singularity respectively. Therefore, it is important to note that these parameters can only describe a black hole under the condition that $r_{-}<r_{+}$. 

The authors have shown for an electrically charged black hole that with this metric, the metric field Eq. (\ref{eq:fRmetricfield}) and Maxwell Eq. (\ref{eq:Maxwell}) are indeed satisfied. To complement this result, we propose to verify whether the solution still holds for a magnetically charged black hole in Sec. \ref{sec:mag}.

\subsection{The $\Lm \rightarrow 0$ limit}\label{sec:LM0}


The on-shell Lagrangian density for the solution in Eqs. (\ref{eq:metric}-\ref{eq:Asol}) reads
\be
\Lm = \frac{Q^2}{(r-r_-)^{8/13} r^{44/13}},
\ee
while at the same time the Ricci scalar is
\be
R = - \frac{Q^2}{(r-r_-)^{6/13} r^{46/13}}.
\ee
In particular, one can check that the ratio $R/\Lm$ is perfectly well defined at $\Lm \rightarrow 0$ that corresponds to $Q \rightarrow 0$:
\be
\vartheta = - \frac{R}{\Lm} = \left(1-\frac{r_-}{r}\right)^{-2/13},
\ee
as already derived in Eq. (\ref{eq:kappa}).

\subsection{Magnetically charged BH}\label{sec:mag}

One can deduce the solution for a black-hole with a magnetic charge by doing the following transformations:
\begin{subequations}\label{eq:transfo}
\bea
&&\vartheta \longrightarrow \vartheta^m = \frac{1}{\vartheta}\textrm{, or }\kappa \longrightarrow \kappa_m = 1/\kappa,\\
&&F_{\mu \nu}  \longrightarrow F^{m}_{\mu \nu} = \frac{1}{2}\vartheta ~ \epsilon_{\mu \nu \kappa \lambda} F^{\kappa \lambda},\\
&&g_{\mu \nu} \longrightarrow g^m_{\mu \nu} = \vartheta^4 g_{\mu \nu},
\eea
\end{subequations}\label{eq:litdiffm}
where $\epsilon_{\mu \nu \kappa \lambda}$ is the Levi-Cività tensor. The resulting field solutions read
\be
ds^2 = -\lambda_{0m}^2 dt^2 + \lambda_{rm}^{-2} dr^2 + \rho_m^2 d\Omega^2,\label{eq:metricm}
\ee
with
\bea
&&\lambda_{0m}^2 = \left(1-\frac{r_+}{r} \right)\kappa_m^{-7/2},\label{eq:NRBH_lambda0m}\\
&&\lambda_{rm}^2 = \left(1-\frac{r_+}{r} \right)\kappa_m^{-15/2},\\
&&\rho_m^2 = r^2 \kappa_m,\label{eq:rhom}
\eea
while the scalar-field solution reads
\be \label{eq:kappam}
\kappa_m = \frac{1}{\sqrt{\phi_m}} = \frac{1}{\vartheta_m}= \left(1-\frac{r_-}{r} \right)^{-2/13}
\ee
and the magnetic field solution is
\be \label{eq:Asolm}
A=-Q \cos{\theta} d\psi
\ee
The verification of the various field equations has been carried out with the open source mathematical language \textit{SageManifolds} \cite{gourgoulhon:2015jc} and can be consulted on Github at the following address: \url{https://github.com/mWavasseur/ER/blob/main/Art.I Slowly rotating and charged BH in ER/Non Rotating BH/ER_NR_Magnetic_BH.ipynb}.

This case is rather interesting as $\Lm$ has its sign changed when going from the electric to the magnetic case. Indeed, the electromagnetic Lagrangian can be written in terms of the electric and magnetic fields as $\Lm \propto E^2 - B^2$, such that $\Lm > 0$ for an electric charge, whereas $\Lm < 0$ for a magnetic one. Despite this change, the ratio $\kappa=R/\Lm$ keeps the same sign, indicating that $R$ also has its sign changed from one case to another. One can verify that explicitely, as $R$ and $\Lm$ read as follows
\bea
R = &Q^2& \frac{\kappa}{r^{52/13}},\\
\Lm = - &Q^2&  \frac{\kappa^{2}}{r^{52/13}}.
\eea
One indeed verifies that $\kappa_m = - R/\Lm=1/\kappa$ where $\kappa$ is given by Eq. (\ref{eq:kappa}).

\subsection{Petrov classification}

The Weyl scalars $\{\Psi_i \}$ are constructed upon the Weyl tensor \cite{griffiths:2009bk}
\bea
C_{\kappa \lambda \mu \nu}&=&R_{\kappa \lambda \mu \nu}-\frac{1}{2}\left(R_{\kappa \mu} g_{\lambda \nu}-R_{\kappa \nu} g_{\lambda \mu}+R_{\lambda \nu} g_{\kappa \mu}-R_{\lambda \mu} g_{\kappa \nu}\right)\nonumber\\
&&+\frac{1}{6}\left(g_{\kappa \mu} g_{\lambda \nu}-g_{\kappa \nu} g_{\lambda \mu}\right) R,
\eea
and the null-tetrads $(l,n,m,\overline{m})$ as follows
\begin{align*}
\Psi_0 &= C_{\kappa \lambda \mu \nu} l^{\kappa} m^\lambda l^{\mu} m^{\nu}, \\
\Psi_1 &= C_{\kappa \lambda \mu \nu} l^{\kappa} n^\lambda l^{\mu} m^{\mu}, \\
\Psi_2 &= C_{\kappa \lambda \mu \nu} l^{\kappa} m^\lambda \overline{m}^{\mu} n^{\nu}, \\
\Psi_3 &= C_{\kappa \lambda \mu \nu} n^{\kappa} l^\lambda n^{\mu} \overline{m}^{\nu}, \\
\Psi_4 &= C_{\kappa \lambda \mu \nu} n^{\kappa} \overline{m}^\lambda n^{\mu} \overline{m}^{\nu},
\end{align*}
where the null-tetrad\footnote{The null complex tetrad can be consulted on Github at the following address: \url{https://github.com/mWavasseur/ER/blob/main/Art.I Slowly rotating and charged BH in ER/Non Rotating BH/ER_NR_Tetrad.ipynb}} has been chosen such that the metric tensor can be written 
\be
g_{\mu \nu} = -k_\mu l_\nu - l_\mu k_\nu + m_\mu \bar m_\nu + \bar m_\mu m_\nu.
\ee

Based on the values of the Weyl scalars $(l,n,m,\overline{m})$, we construct the Petrov classification which indicates the number of principal null directions that spacetime has and their multiplicity. This classification allows us to verify if the properties of spacetime within the framework of Entangled Relativity exhibit the same algebraic classification as in the framework of General Relativity when describing black holes with or without rotation.

We have calculated the Weyl scalars for the cases of a black hole charged electrically on one hand, and magnetically on the other. In each case, the number and multiplicity of spacetime directions correspond to a \textit{type D} black hole. Thus, we obtain the same algebraic classification as in General Relativity (e.g., Schwarzschild and Kerr) for which there are two distinct and orthogonal principal null directions. The magnetic and electric cases have the same classification since their metric solutions only differ by a conformal factor in Eq. (\ref{eq:transfo}), while the Petrov classification is invariant under conformal transformations. The Petrov classification\footnote{Note that the notebook for the determination of the complex null tetrad must be executed beforehand.} can be consulted on Github at the following address: \url{https://github.com/mWavasseur/ER/blob/main/Art.I Slowly rotating and charged BH in ER/Non Rotating BH/ER_NR_Petrov_Classification.ipynb}

\section{Slowly rotating charged BH}

Here we propose to extend the work of \cite{minazzoli:2021ej} by exploring the case of slowly rotating black holes. In order to derive the solution to the field equations in the slowly rotating case, we adopt the same strategy as in \cite{minazzoli:2021ej} for a non rotating black hole: we deduce the solution of Entangled Relativity from the conformal transformation of the solution of the Einstein-Maxwell-dilaton theory found in \cite{horne:1992pr}, and with the specific value for its coupling parameter that corresponds to Entangled Relativity \cite{minazzoli:2021ej}.\\

By a `\textit{slowly rotating black-hole}', we mean that we check that the fields are solution to the equations to first order in the angular momentum parameter $a$. Finding a more general analytical solution has proven to be very difficult in Einstein-Maxwell-dilaton theories in general \cite{hirschmann:2018pr}, except for the cases that correspond to a Kaluza-Klein theory with compactified dimensions \cite{horne:1992pr}. The solution, now axially symmetric instead of spherically symmetric, can be expressed as:

\begin{equation}
\label{eq:metricSRBHe}
    ds^{2}=-\lambda_{0}^{2}dt^{2}+\lambda_{r}^{-2}dr^{2}+\rho^{2}d\Omega^2-2af(r)\sin{\theta}^{2}dtd\psi,
\end{equation}

where
\bea
&&\lambda_0^2 = \left(1-\frac{r_+}{r} \right)\kappa^{15/2},\label{eq:RBGH_lambda0}\\
&&\lambda_r^2 = \left(1-\frac{r_+}{r} \right)\kappa^{7/2},\\
&&\rho^2 = r^2 \kappa^{3},\label{eq:rhor}\\
&&\kappa = \left(1-\frac{r_-}{r}\right)^{2/13},
\eea

The non diagonal term is driven by the function $f(r)$:

\begin{equation}
\label{eq:f(r)}
    f(r)=\frac{169 r^2 \left(1-\frac{r_{-}}{r}\right)^{\frac{2}{13}}}{99 r_{-}^2} - \left(1-\frac{r_{-}}{r}\right)^{\frac{11}{13}} \left( 1 + \frac{169 r^2}{99 r_{-}^2} + \frac{13r}{11 r_{-}} - \frac{r_{+}}{r} \right) ,
\end{equation}
which is nonlinear in $r$ and has been designed \cite{horne:1992pr} such that\\ $f(r) \underset{r\rightarrow\infty}{=}\frac{35r_{-}+39r_{+}}{39r} \propto1/r$.\\

The dilaton remains unaffected as do the expressions for the mass and charge, which are still given by Eq. (\ref{eq:masse-charge}), but the $\Psi$-component of the vector potential becomes nonzero:

\begin{equation}
\label{eq:potential}
A_{t} = -\frac{Q}{r}, \hspace{0.5cm}
A_{\psi} = a\sin^2{\theta}\frac{Q}{r}.
\end{equation}

We note above all that the expression Eq. (\ref{eq:metricSRBHe}) only differs from the non-rotating case (\cite{minazzoli:2021ej}, \cite{horne:1992pr}) by a non-diagonal term, similarly to the solutions of Kerr-Newman and rotating charged Kaluza-Klein \cite{horne:1992pr}. 

It is interesting to note that the on-shell value of the matter Lagrangian reduces to
\be
\Lm = \frac{Q^2}{(\kappa r)^4} + \Ol(a^2),
\ee
whereas the Ricci scalar is
\be
R = -\frac{Q^2}{\kappa^3 r^4} + \Ol(a^2),
\ee
such that one indeed recovers Eq. (\ref{eq:defkappa}) at leading order in $a$: $-R/\Lm = \kappa+ \Ol(a^2)$.

Given the extensive complexity of the calculations to verify if Eq. (\ref{eq:metricSRBHe}) is indeed a solution to Eq. (\ref{eq:fRmetricfield}) and Eq. (\ref{eq:Maxwell}), we use the open source mathematical language \textit{SageManifolds} \cite{gourgoulhon:2015jc}, which is based on \textit{SageMath} \cite{stein:2005ab} to reinject the metric described by Eqs. (\ref{eq:metricSRBHe}-\ref{eq:rhor}) into the field equation (\ref{eq:fRmetricfield}). The result is indeed verified, and therefore, Eq. (\ref{eq:metricSRBHe}) is a solution in Entangled Relativity for a slowly rotating charged black hole. (See notebook\footnote{Note that the notebook for the determination of the complex null tetrad, provided in Sec. \ref{sec:petrov}, must be executed beforehand.} on Github at the following address: \url{https://github.com/mWavasseur/ER/blob/main/Art.I Slowly rotating and charged BH in ER/Slowly Rotating BH/ER_SR_Electric_BH.ipynb}.) 

In the notebook, in order to maintain the various calculations of the rotating case in a tractable way, we simplify the numerical processing by performing intermediate first-order Taylor expansions. This choice does not compromise the validity of our results, as they are part of a first-order analysis in `a'.

\subsection{Near vacuum limit and Kerr solution}

Now let's see how the classical Kerr solution for rotating black holes in General Relativity can be interpreted as good approximations of the solutions of Entangled Relativity in case of a slow rotation and when the charge goes to zero. Starting from the Kerr-Newman solution, we perform a simple first-order Taylor expansion in $a$ of the metric components to approximate the spacetime for slow rotations. We get the following expression:

\begin{eqnarray}
\label{eq:KerrNewman}
    ds^{2} &=& -\nu^{2}\mathrm{d} t^{2} + \nu^{-2}\mathrm{d} r^{2}  + r^{2} \mathrm{d} {\Omega^2} - 2h(r)a \sin\left({\theta}\right)^{2} \mathrm{d} t\mathrm{d} {\phi},
\end{eqnarray}

\noindent where $\nu^{2}=\frac{Q^{2} - 2 \, M r + r^{2}}{r^{2}}$ and $h(r)=\frac{2M r - Q^{2}}{r^{2}}$ with $M,Q$ the mass and charge. The usual quantities  $\Sigma$ and  $\Delta$ simply reduce to $r^{2}$  and $r^{2}+Q^{2}-2Mr$ respectively. This form is obviously very similar to the solution Eq. (\ref{eq:metricSRBHe}). 

\indent Let's now look at how the solution in Entangled Relativity behaves in the near-vacuum limit $Q\to 0$ to see if it converges towards the predictions of General Relativity. As we mentioned above, we logically compare it with the slowly rotating Kerr-Newman solution Eq. (\ref{eq:KerrNewman}). By reinjecting the expressions of the charge and the mass, given by Eq. (\ref{eq:Mass}) and Eq. (\ref{eq: Charge}) respectively, into the metric Eq. (\ref{eq:KerrNewman}) we get the following limits:

\begin{eqnarray}
\label{eq:limits}
    \lambda_{0}^{2}-\nu^{2}&= &\frac{3 \, r_{+} - 4 \, r }{13 \, r^{2}}~ r_-+ \mathcal{O}(r_{-}^{2}), \nonumber\\
    f(r)-h(r)&=&\frac{{2 \, a r  - 9 \, a r_{+}}}{39 \, r^{2}}~ r_-+ \mathcal{O}(r_{-}^{2}),\nonumber\\
    \rho(r)-r^{2}&=&\frac{6}{13} \, r r_{-}+ \mathcal{O}(r_{-}^{2}),\nonumber\\    
    \lambda^{-2}-\nu^{-2}&=&\frac{r(52r - 65r_{+})+r_{-}(77r-84r_{+})}{13\left(r - r_{+}\right)(13r(r-r_{+})+r_{-}(12r_{+}-11r))}~ r_-+ \mathcal{O}(r_{-}^{2}).\nonumber
\end{eqnarray}

 We find that to first order in $a$, the difference between the Kerr-Newman metric components and those of Entangled Relativity is $\mathcal{O}(r_{-})$. Therefore, slowly rotating Kerr-Newman is not solution of Entangled Relativity. However, we recall that the limit as $r_{-}\to0$ corresponds to the absence of matter fields $Q=0$ or $T_{\mu\nu}\sim0$, thereby describing a slowly rotating uncharged black hole. Thus, the external metric of the Kerr Solution (in the case of a slow rotation) constitutes a very good approximation---in a near-vacuum situation---of the solution of a slowly rotating charged black hole in Entangled Relativity. This was anticipated in \cite{minazzoli:2021ej}.

Nevertheless, there exists a branch with scalar hair ($r_+ \to 0$ while $r_- \neq 0$), which does not represent black holes per se but rather naked singularities. Although this branch exists, it is generally assumed that its solutions cannot form through the collapse of a star, as scalar hair is typically radiated away via monopolar gravitational waves during the collapse of a star into a black hole—see \cite{gerosa:2016cq} and references therein.

 (See notebook on Github at the following address: \url{https://github.com/mWavasseur/ER/blob/main/Art.I Slowly rotating and charged BH in ER/Slowly Rotating BH/ER_SR_Electric_BH.ipynb}.).

\subsection{Magnetically charged BH}

Using the same transformations as those described in section \ref{sec:mag}, we obtain in the case of slow rotation the solution for a black hole with a magnetic charge. Following the strategy mentioned before, we verify that the external solution ($r>r_{+})$ perfectly satisfies the various field equations in the notebook available on Github at the following address: \url{https://github.com/mWavasseur/ER/blob/main/Art.I Slowly rotating and charged BH in ER/Slowly Rotating BH/ER_SR_Magnetic_BH.ipynb}.

The field solutions read for a magnetically charged BH:
\begin{equation}
\label{eq:metricSRBHm}
    ds^{2}=-\lambda_{0m}^{2}dt^{2}+\lambda_{rm}^{-2}dr^{2}+\rho_{m}^{2}d\Omega^2-2af_{m}(r)\sin{\theta}^{2}dtd\psi,
\end{equation}

where
\bea
&&\lambda_{0m}^2 = \left(1-\frac{r_+}{r} \right)\kappa_m^{-7/2},\label{eq:NRBH_lambda0mr}\\
&&\lambda_{rm}^2 = \left(1-\frac{r_+}{r} \right)\kappa_m^{-15/2},\\
&&\rho_m^2 = r^2 \kappa_m,\label{eq:rhomr}\\
&&f_{m}(r) = \kappa_{m}^{-4}f(r)
\eea
where the function $f(r)$ is from (\ref{eq:f(r)}) and $\kappa_{m}$ is given in (\ref{eq:kappam}). The vector potential reads:
\begin{equation}
\label{eq:potentialm}
A_{t} = \frac{13 Q a \cos(\theta)}{99 r_{-}^2} \left(\left(13 +9\frac{r_{-}}{r}\right)\left(\frac{r - r_{-}}{r}\right)^{\frac{9}{13}} - 13 \right)
, \hspace{0.5cm}
A_{\psi} = Q \cos(\theta).
\end{equation}

The on-shell value of the matter Lagrangian reduces to
\be
\Lm = -\frac{Q^2}{\kappa_{m}^2r^4} + \Ol(a^2),
\ee
whereas the Ricci scalar is
\be
R = \frac{Q^2}{\kappa_{m} r^4} + \Ol(a^2),
\ee
such that one indeed recovers Eq. (\ref{eq:defkappa}) at leading order in $a$: $-R/\Lm = \kappa_{m} + \Ol(a^2)$.

\subsection{Choice of a null complex tetrad}

Additionally, we have constructed a valid tetrad within the framework of Entangled Relativity for the case of slow rotation. We proceed step by step, starting with a tetrad that is valid for the Kerr metric, we verify that the conditions still hold in the case of slow rotation (first order in $a$). Then, we gradually simplify it while ensuring that the tetrad's orthogonality remain intact. Ultimately, we achieve a fairly general tetrad form for a non-diagonal metric that is applicable in cases of slow rotation. We get that the following family of vectors

\bea
l^{\mu} &=& \left[-\frac{1}{\sigma\left(r\right)}, 1, 0, -\frac{a j\left(r\right)}{{\left(\sigma\left(r\right) + 1\right)} \epsilon\left(r\right) \sigma\left(r\right)}\right],\\ \label{eq:KerrTetrad}
n^{\mu} &=&  \left[\frac{1}{2}, \frac{1}{2} \, \sigma\left(r\right), 0, \frac{a j\left(r\right)}{2 \, {\left(\sigma\left(r\right) + 1\right)} \epsilon\left(r\right)}\right],\nonumber\\
m^{\mu} &=& \left[\frac{i \, \sqrt{2} a j\left(r\right) \sin\left({\theta}\right)}{2 \, {\left(\sigma\left(r\right) + 1\right)} \sqrt{\epsilon\left(r\right)}}, 0, \frac{\sqrt{2}}{2 \, \sqrt{\epsilon\left(r\right)}}, \frac{i \, \sqrt{2}}{2 \, \sqrt{\epsilon\left(r\right)} \sin\left({\theta}\right)}\right],\nonumber\\
\bar{m}^{\mu} &=&  \left[-\frac{i \, \sqrt{2} a j\left(r\right) \sin\left({\theta}\right)}{2 \, {\left(\sigma\left(r\right) + 1\right)} \sqrt{\epsilon\left(r\right)}}, 0, \frac{\sqrt{2}}{2 \, \sqrt{\epsilon\left(r\right)}}, -\frac{i \, \sqrt{2}}{2 \, \sqrt{\epsilon\left(r\right)} \sin\left({\theta}\right)}\right],\nonumber
\eea
defines a complex null tetrad for all non-diagonal metrics of the form

\begin{eqnarray}
    ds^{2} &=& \sigma^{2}(r)\mathrm{d} t^{2} - \sigma^{-2}(r)\mathrm{d} r^{2}  + \epsilon^{2}(r) \mathrm{d} {\Omega^2} - 2j(r)a \sin\left({\theta}\right)^{2} \mathrm{d} t\mathrm{d} {\psi}.
\end{eqnarray}
where $\sigma, \epsilon$ and $j$ are arbitrary functions of $r$.

The different steps for constructing this null complex tetrad can be found in the notebook available at the following address: \url{https://github.com/mWavasseur/ER/blob/main/Sage_notebooks/ER_SR_Null_Tetrad.ipynb}.

\subsection{Petrov classification}
\label{sec:petrov}

As we did for the non-rotating case, we have calculated the Weyl scalars for a black hole that is electrically charged on one hand, and magnetically on the other hand. It results that the number and multiplicity of the directions escape the Petrov classification as both the scalars $\Psi_{1}$ and $\Psi_{3}$ become non zero. However, we notice that these two scalars are now purely imaginary numbers which reduce to zero in the nearly empty limit $r_- \rightarrow 0$. Thus, we recover the symmetries of a type D spacetime we found for the non rotating case. To confirm these results, we also verify that in this limit, the scalar $\Psi_{2}$ indeed has the expression obtained for the Kerr metric, since as we saw previously, the slowly rotating Kerr solution represents a good approximation of the Entangled Relativity solution in the case where $r_{-}$ tends to zero. The Petrov classification has been carried out and can be consulted in the notebook\footnote{The notebook is quite heavy and may sometimes cause crashes upon opening. In such cases, we recommend clearing the outputs using the terminal command:  \lstinline|jupyter nbconvert --clear-output --inplace your_notebook.ipynb|} available at the following address: \url{https://github.com/mWavasseur/ER/blob/main/Art.I Slowly rotating and charged BH in ER/Slowly Rotating BH/ER_SR_Petrov_Classification.ipynb}..

\section{Conclusion}

We have considered here the weakly rotating case of a charged black hole in continuity with \cite{minazzoli:2021ej}. Based on previous studies, notably the results in \cite{horne:1992pr}, we tested if solutions developed within the framework of the Maxwell-dilaton theory indeed satisfy the equations of Entangled Relativity for both an electrically and a magnetically charged black-hole. Delving into the case of Entangled Relativity is particularly interesting because, in this theory, the dilaton field is defined as the ratio between $R$ and $\Lm$. Therefore, the limit $\Lm \rightarrow 0$ is of special interest in this context, compared to other Einstein-Maxwell-dilaton theories, such as bosonic strings or Kaluza-Klein theories \cite{horne:1992pr}. Notably, we have once again verified \cite{minazzoli:2021ej,minazzoli:2024ar} that $R/\Lm$ remains well defined in the $\Lm \rightarrow 0$ limit.

Thus, in (\ref{eq:metricSRBHe}-\ref{eq:potential}) and in (\ref{eq:metricSRBHm}-\ref{eq:potentialm}), we present an axisymmetric solution valid in the simple case of a matter field with infinite range composed by electromagnetism only, and with the assumption that the black-hole is slowly rotating. In the nearly vacuum limit $T_{\mu\nu}\sim0$, the solution reduces to the usual form of the slowly rotating Kerr solution, as \cite{minazzoli:2021ej} anticipated.\\

Finding an analytical solution valid for all amplitude of rotation may prove to be difficult, the same way that it has been, as far as we know, so far impossible to find a general rotating solution for Einstein-Maxwell-dilaton theories \cite{horne:1992pr}.\\

We have also classified spacetimes for the non-rotating and slowly-rotating cases. In the first case, we find a number and multiplicity of null directions corresponding to type D, whereas in the second case, the structure escapes the Petrov classification criteria. However, we note that the deviation from these criteria remains small and that it returns to them in the near-vacuum limit.\\

These results seem to confirm that one should not expect large deviations from Entangled Relativity with respect to General Relativity in black-hole physics---see also \cite{minazzoli:2024ar}. Compact objects such as neutron stars or white dwarfs may be better targets to constrain Entangled Relativity \cite{arruga:2021pr,arruga:2021ep}.

\bmhead{Acknowledgments}

We sincerely thank Eric Gourgoulhon  for his invaluable guidance in optimizing our notebooks using the open-source mathematical software SageManifolds.

\bibliography{ER_slowR}


\begin{thebibliography}{30}
\ifx \bisbn   \undefined \def \bisbn  #1{ISBN #1}\fi
\ifx \binits  \undefined \def \binits#1{#1}\fi
\ifx \bauthor  \undefined \def \bauthor#1{#1}\fi
\ifx \batitle  \undefined \def \batitle#1{#1}\fi
\ifx \bjtitle  \undefined \def \bjtitle#1{#1}\fi
\ifx \bvolume  \undefined \def \bvolume#1{\textbf{#1}}\fi
\ifx \byear  \undefined \def \byear#1{#1}\fi
\ifx \bissue  \undefined \def \bissue#1{#1}\fi
\ifx \bfpage  \undefined \def \bfpage#1{#1}\fi
\ifx \blpage  \undefined \def \blpage #1{#1}\fi
\ifx \burl  \undefined \def \burl#1{\textsf{#1}}\fi
\ifx \doiurl  \undefined \def \doiurl#1{\url{https://doi.org/#1}}\fi
\ifx \betal  \undefined \def \betal{\textit{et al.}}\fi
\ifx \binstitute  \undefined \def \binstitute#1{#1}\fi
\ifx \binstitutionaled  \undefined \def \binstitutionaled#1{#1}\fi
\ifx \bctitle  \undefined \def \bctitle#1{#1}\fi
\ifx \beditor  \undefined \def \beditor#1{#1}\fi
\ifx \bpublisher  \undefined \def \bpublisher#1{#1}\fi
\ifx \bbtitle  \undefined \def \bbtitle#1{#1}\fi
\ifx \bedition  \undefined \def \bedition#1{#1}\fi
\ifx \bseriesno  \undefined \def \bseriesno#1{#1}\fi
\ifx \blocation  \undefined \def \blocation#1{#1}\fi
\ifx \bsertitle  \undefined \def \bsertitle#1{#1}\fi
\ifx \bsnm \undefined \def \bsnm#1{#1}\fi
\ifx \bsuffix \undefined \def \bsuffix#1{#1}\fi
\ifx \bparticle \undefined \def \bparticle#1{#1}\fi
\ifx \barticle \undefined \def \barticle#1{#1}\fi
\bibcommenthead
\ifx \bconfdate \undefined \def \bconfdate #1{#1}\fi
\ifx \botherref \undefined \def \botherref #1{#1}\fi
\ifx \url \undefined \def \url#1{\textsf{#1}}\fi
\ifx \bchapter \undefined \def \bchapter#1{#1}\fi
\ifx \bbook \undefined \def \bbook#1{#1}\fi
\ifx \bcomment \undefined \def \bcomment#1{#1}\fi
\ifx \oauthor \undefined \def \oauthor#1{#1}\fi
\ifx \citeauthoryear \undefined \def \citeauthoryear#1{#1}\fi
\ifx \endbibitem  \undefined \def \endbibitem {}\fi
\ifx \bconflocation  \undefined \def \bconflocation#1{#1}\fi
\ifx \arxivurl  \undefined \def \arxivurl#1{\textsf{#1}}\fi
\csname PreBibitemsHook\endcsname

\bibitem[\protect\citeauthoryear{{Ludwig} et~al.}{2015}]{ludwig:2015pl}
\begin{barticle}
\bauthor{\bsnm{{Ludwig}}, \binits{H.}},
\bauthor{\bsnm{{Minazzoli}}, \binits{O.}},
\bauthor{\bsnm{{Capozziello}}, \binits{S.}}:
\batitle{{Merging matter and geometry in the same Lagrangian}}.
\bjtitle{Physics Letters B}
\bvolume{751},
\bfpage{576}--\blpage{578}
(\byear{2015})
\doiurl{10.1016/j.physletb.2015.11.023}
{\href{https://arxiv.org/abs/1506.03278}{{arXiv:1506.03278}}}
{[gr-qc]}
\end{barticle}
\endbibitem

\bibitem[\protect\citeauthoryear{{Arruga} et~al.}{2021}]{arruga:2021pr}
\begin{barticle}
\bauthor{\bsnm{{Arruga}}, \binits{D.}},
\bauthor{\bsnm{{Rousselle}}, \binits{O.}},
\bauthor{\bsnm{{Minazzoli}}, \binits{O.}}:
\batitle{{Compact objects in entangled relativity}}.
\bjtitle{\prd}
\bvolume{103}(\bissue{2}),
\bfpage{024034}
(\byear{2021})
\doiurl{10.1103/PhysRevD.103.024034}
{\href{https://arxiv.org/abs/2011.14629}{{arXiv:2011.14629}}}
{[gr-qc]}
\end{barticle}
\endbibitem

\bibitem[\protect\citeauthoryear{{Einstein}}{1918}]{einstein:1918an}
\begin{barticle}
\bauthor{\bsnm{{Einstein}}, \binits{A.}}:
\batitle{{Prinzipielles zur allgemeinen Relativitatstheorie}}.
\bjtitle{Annalen der Physik}
\bvolume{360}(\bissue{4}),
\bfpage{241}--\blpage{244}
(\byear{1918})
\doiurl{10.1002/andp.19183600402} .
\bcomment{Translation available at
  \url{https://einsteinpapers.press.princeton.edu/vol7-trans/49}}
\end{barticle}
\endbibitem

\bibitem[\protect\citeauthoryear{{Einstein}}{1917}]{einstein:1917co}
\begin{botherref}
\oauthor{\bsnm{{Einstein}}, \binits{A.}}:
{Kosmologische Betrachtungen zur allgemeinen Relativit{\"a}tstheorie}.
Sitzungsberichte der K{\"o}niglich Preu{\ss}ischen Akademie der Wissenschaften
  (Berlin,
142--152
(1917).
Translation available at
  \url{https://einsteinpapers.press.princeton.edu/vol6-trans/433}
\end{botherref}
\endbibitem

\bibitem[\protect\citeauthoryear{{Einstein}}{1918}]{einstein:1918sp}
\begin{botherref}
\oauthor{\bsnm{{Einstein}}, \binits{A.}}:
{Kritisches zu einer von Hrn. de Sitter gegebenen L{\"o}sung der
  Gravitationsgleichungen}.
Sitzungsberichte der K{\"o}niglich Preu{\ss}ischen Akademie der Wissenschaften
  (Berlin),
270--272
(1918).
Translation available at
  \url{https://einsteinpapers.press.princeton.edu/vol7-trans/52}
\end{botherref}
\endbibitem

\bibitem[\protect\citeauthoryear{{Hoefer}}{1995}]{hoefer:1995cf}
\begin{bchapter}
\bauthor{\bsnm{{Hoefer}}, \binits{C.}}:
\bctitle{{Einstein's Formulations of Mach's Principle}}.
In: \beditor{\bsnm{{Barbour}}, \binits{J.B.}},
\beditor{\bsnm{{Pfister}}, \binits{H.}} (eds.)
\bbtitle{Mach's Principle: From Newton's Bucket to Quantum Gravity},
p. \bfpage{67}.
\bpublisher{{Birkh\"aser}},
\blocation{Boston University}
(\byear{1995})
\end{bchapter}
\endbibitem

\bibitem[\protect\citeauthoryear{{Pais}}{1982}]{pais:1982bk}
\begin{bbook}
\bauthor{\bsnm{{Pais}}, \binits{A.}}:
\bbtitle{{Subtle Is the Lord. The Science and the Life of Albert Einstein}}.
\bpublisher{Oxford University Press},
\blocation{Oxford}
(\byear{1982})
\end{bbook}
\endbibitem

\bibitem[\protect\citeauthoryear{{Minazzoli}}{2024}]{minazzoli:2024pn}
\begin{barticle}
\bauthor{\bsnm{{Minazzoli}}, \binits{O.}}:
\batitle{{On the Principle of Relativity of Inertia in Both General and
  Entangled Relativities}}.
\bjtitle{Physics of Particles and Nuclei}
\bvolume{55}(\bissue{6}),
\bfpage{1488}--\blpage{1493}
(\byear{2024})
\doiurl{10.1134/S1063779624701132}
\end{barticle}
\endbibitem

\bibitem[\protect\citeauthoryear{{Minazzoli}}{2022}]{minazzoli:2022ar}
\begin{botherref}
\oauthor{\bsnm{{Minazzoli}}, \binits{O.}}:
{Quantum of action in entangled relativity}.
arXiv e-prints,
2206--03824
(2022)
\doiurl{10.48550/arXiv.2206.03824}
{\href{https://arxiv.org/abs/2206.03824}{{arXiv:2206.03824}}}
{[gr-qc]}
\end{botherref}
\endbibitem

\bibitem[\protect\citeauthoryear{{Minazzoli}}{2023}]{minazzoli:2023ar}
\begin{botherref}
\oauthor{\bsnm{{Minazzoli}}, \binits{O.}}:
{Standard quantum field theory from entangled relativity}.
Contribution to the 2023 Gravitation session of the 57th Rencontres de Moriond,
2304--09482
(2023)
\doiurl{10.48550/arXiv.2304.09482}
{\href{https://arxiv.org/abs/2304.09482}{{arXiv:2304.09482}}}
{[gr-qc]}
\end{botherref}
\endbibitem

\bibitem[\protect\citeauthoryear{{Misner} et~al.}{1973}]{MTW}
\begin{bbook}
\bauthor{\bsnm{{Misner}}, \binits{C.W.}},
\bauthor{\bsnm{{Thorne}}, \binits{K.S.}},
\bauthor{\bsnm{{Wheeler}}, \binits{J.A.}}:
\bbtitle{{Gravitation}},
(\byear{1973})
\end{bbook}
\endbibitem

\bibitem[\protect\citeauthoryear{{Minazzoli} and
  {Hees}}{2013}]{minazzoli:2013pr}
\begin{barticle}
\bauthor{\bsnm{{Minazzoli}}, \binits{O.}},
\bauthor{\bsnm{{Hees}}, \binits{A.}}:
\batitle{{Intrinsic Solar System decoupling of a scalar-tensor theory with a
  universal coupling between the scalar field and the matter Lagrangian}}.
\bjtitle{\prd}
\bvolume{88}(\bissue{4}),
\bfpage{041504}
(\byear{2013})
\doiurl{10.1103/PhysRevD.88.041504}
{\href{https://arxiv.org/abs/1308.2770}{{arXiv:1308.2770}}}
{[gr-qc]}
\end{barticle}
\endbibitem

\bibitem[\protect\citeauthoryear{{LIGO Scientific Collaboration} and {Virgo
  Collaboration}}{2016}]{GW150914}
\begin{barticle}
\bauthor{\bsnm{{LIGO Scientific Collaboration}}},
\bauthor{\bsnm{{Virgo Collaboration}}}:
\batitle{{Observation of Gravitational Waves from a Binary Black Hole Merger}}.
\bjtitle{\prl}
\bvolume{116}(\bissue{6}),
\bfpage{061102}
(\byear{2016})
\doiurl{10.1103/PhysRevLett.116.061102}
{\href{https://arxiv.org/abs/1602.03837}{{arXiv:1602.03837}}}
{[gr-qc]}
\end{barticle}
\endbibitem

\bibitem[\protect\citeauthoryear{{Event Horizon Telescope
  Collaboration}}{2019}]{EHT:2019aj}
\begin{barticle}
\bauthor{\bsnm{{Event Horizon Telescope Collaboration}}}:
\batitle{{First M87 Event Horizon Telescope Results. I. The Shadow of the
  Supermassive Black Hole}}.
\bjtitle{\apjl}
\bvolume{875}(\bissue{1}),
\bfpage{1}
(\byear{2019})
\doiurl{10.3847/2041-8213/ab0ec7}
{\href{https://arxiv.org/abs/1906.11238}{{arXiv:1906.11238}}}
{[astro-ph.GA]}
\end{barticle}
\endbibitem

\bibitem[\protect\citeauthoryear{{Minazzoli} and
  {Santos}}{2021}]{minazzoli:2021ej}
\begin{barticle}
\bauthor{\bsnm{{Minazzoli}}, \binits{O.}},
\bauthor{\bsnm{{Santos}}, \binits{E.}}:
\batitle{{Charged black hole and radiating solutions in entangled relativity}}.
\bjtitle{European Physical Journal C}
\bvolume{81}(\bissue{7}),
\bfpage{640}
(\byear{2021})
\doiurl{10.1140/epjc/s10052-021-09441-w}
{\href{https://arxiv.org/abs/2102.10541}{{arXiv:2102.10541}}}
{[gr-qc]}
\end{barticle}
\endbibitem

\bibitem[\protect\citeauthoryear{{Harko} and {Lobo}}{2010}]{harko:2010ep}
\begin{barticle}
\bauthor{\bsnm{{Harko}}, \binits{T.}},
\bauthor{\bsnm{{Lobo}}, \binits{F.S.N.}}:
\batitle{{f( R, L $_{ m }$) gravity}}.
\bjtitle{European Physical Journal C}
\bvolume{70}(\bissue{1-2}),
\bfpage{373}--\blpage{379}
(\byear{2010})
\doiurl{10.1140/epjc/s10052-010-1467-3}
{\href{https://arxiv.org/abs/1008.4193}{{arXiv:1008.4193}}}
{[gr-qc]}
\end{barticle}
\endbibitem

\bibitem[\protect\citeauthoryear{{Arruga} and
  {Minazzoli}}{2021}]{arruga:2021ep}
\begin{barticle}
\bauthor{\bsnm{{Arruga}}, \binits{D.}},
\bauthor{\bsnm{{Minazzoli}}, \binits{O.}}:
\batitle{{Analytical external spherical solutions in entangled relativity}}.
\bjtitle{European Physical Journal C}
\bvolume{81}(\bissue{11}),
\bfpage{1027}
(\byear{2021})
\doiurl{10.1140/epjc/s10052-021-09818-x}
{\href{https://arxiv.org/abs/damour2106.03426}{{arXiv:damour2106.03426}}}
{[gr-qc]}
\end{barticle}
\endbibitem

\bibitem[\protect\citeauthoryear{{Minazzoli}}{2018}]{minazzoli:2018pr}
\begin{barticle}
\bauthor{\bsnm{{Minazzoli}}, \binits{O.}}:
\batitle{{Rethinking the link between matter and geometry}}.
\bjtitle{\prd}
\bvolume{98}(\bissue{12}),
\bfpage{124020}
(\byear{2018})
\doiurl{10.1103/PhysRevD.98.124020}
{\href{https://arxiv.org/abs/1811.05845}{{arXiv:1811.05845}}}
{[gr-qc]}
\end{barticle}
\endbibitem

\bibitem[\protect\citeauthoryear{Capozziello and
  Laurentis}{2015}]{capozziello:2015sc}
\begin{barticle}
\bauthor{\bsnm{Capozziello}, \binits{S.}},
\bauthor{\bsnm{Laurentis}, \binits{M.D.}}:
\batitle{{F}({R}) theories of gravitation}.
\bjtitle{Scholarpedia}
\bvolume{10}(\bissue{2}),
\bfpage{31422}
(\byear{2015})
\doiurl{10.4249/scholarpedia.31422} .
\bcomment{revision \#147843}
\end{barticle}
\endbibitem

\bibitem[\protect\citeauthoryear{{Teyssandier} and
  {Tourrenc}}{1983}]{teyssandier:1983jm}
\begin{barticle}
\bauthor{\bsnm{{Teyssandier}}, \binits{P.}},
\bauthor{\bsnm{{Tourrenc}}, \binits{P.}}:
\batitle{{The Cauchy problem for the R+R$^{2}$ theories of gravity without
  torsion}}.
\bjtitle{Journal of Mathematical Physics}
\bvolume{24}(\bissue{12}),
\bfpage{2793}--\blpage{2799}
(\byear{1983})
\doiurl{10.1063/1.525659}
\end{barticle}
\endbibitem

\bibitem[\protect\citeauthoryear{{Jakubiec} and
  {Kijowski}}{1988}]{jakubiec:1988pr}
\begin{barticle}
\bauthor{\bsnm{{Jakubiec}}, \binits{A.}},
\bauthor{\bsnm{{Kijowski}}, \binits{J.}}:
\batitle{{On theories of gravitation with nonlinear Lagrangians}}.
\bjtitle{\prd}
\bvolume{37}(\bissue{6}),
\bfpage{1406}--\blpage{1409}
(\byear{1988})
\doiurl{10.1103/PhysRevD.37.1406}
\end{barticle}
\endbibitem

\bibitem[\protect\citeauthoryear{{Woodard}}{2007}]{woodard:2007ln}
\begin{bchapter}
\bauthor{\bsnm{{Woodard}}, \binits{R.}}:
\bctitle{{Avoiding Dark Energy with 1/R Modifications of Gravity}}.
In: \beditor{\bsnm{{Papantonopoulos}}, \binits{L.}} (ed.)
\bbtitle{The Invisible Universe: Dark Matter and Dark Energy}
vol. \bseriesno{720},
p. \bfpage{403}
(\byear{2007}).
\doiurl{10.1007/978-3-540-71013-4_14}
\end{bchapter}
\endbibitem

\bibitem[\protect\citeauthoryear{{Minazzoli} and
  {Hees}}{2014}]{minazzoli:2014pr}
\begin{barticle}
\bauthor{\bsnm{{Minazzoli}}, \binits{O.}},
\bauthor{\bsnm{{Hees}}, \binits{A.}}:
\batitle{{Late-time cosmology of a scalar-tensor theory with a universal
  multiplicative coupling between the scalar field and the matter Lagrangian}}.
\bjtitle{\prd}
\bvolume{90}(\bissue{2}),
\bfpage{023017}
(\byear{2014})
\doiurl{10.1103/PhysRevD.90.023017}
{\href{https://arxiv.org/abs/1404.4266}{{arXiv:1404.4266}}}
{[gr-qc]}
\end{barticle}
\endbibitem

\bibitem[\protect\citeauthoryear{{Gourgoulhon}
  et~al.}{2015}]{gourgoulhon:2015jc}
\begin{bchapter}
\bauthor{\bsnm{{Gourgoulhon}}, \binits{E.}},
\bauthor{\bsnm{{Bejger}}, \binits{M.}},
\bauthor{\bsnm{{Mancini}}, \binits{M.}}:
\bctitle{{Tensor calculus with open-source software: the SageManifolds
  project}}.
In: \bbtitle{Journal of Physics Conference Series}.
\bsertitle{Journal of Physics Conference Series},
vol. \bseriesno{600},
p. \bfpage{012002}.
\bpublisher{IOP}, \blocation{???}
(\byear{2015}).
\doiurl{10.1088/1742-6596/600/1/012002}
\end{bchapter}
\endbibitem

\bibitem[\protect\citeauthoryear{{Griffiths} and
  {Podolsk{\'y}}}{2009}]{griffiths:2009bk}
\begin{bbook}
\bauthor{\bsnm{{Griffiths}}, \binits{J.B.}},
\bauthor{\bsnm{{Podolsk{\'y}}}, \binits{J.}}:
\bbtitle{{Exact Space-Times in Einstein's General Relativity}}.
\bpublisher{Cambridge University Press}, \blocation{???}
(\byear{2009})
\end{bbook}
\endbibitem

\bibitem[\protect\citeauthoryear{{Horne} and {Horowitz}}{1992}]{horne:1992pr}
\begin{barticle}
\bauthor{\bsnm{{Horne}}, \binits{J.H.}},
\bauthor{\bsnm{{Horowitz}}, \binits{G.T.}}:
\batitle{{Rotating dilaton black holes}}.
\bjtitle{\prd}
\bvolume{46}(\bissue{4}),
\bfpage{1340}--\blpage{1346}
(\byear{1992})
\doiurl{10.1103/PhysRevD.46.1340}
{\href{https://arxiv.org/abs/hep-th/9203083}{{arXiv:hep-th/9203083}}}
{[hep-th]}
\end{barticle}
\endbibitem

\bibitem[\protect\citeauthoryear{{Hirschmann} et~al.}{2018}]{hirschmann:2018pr}
\begin{barticle}
\bauthor{\bsnm{{Hirschmann}}, \binits{E.W.}},
\bauthor{\bsnm{{Lehner}}, \binits{L.}},
\bauthor{\bsnm{{Liebling}}, \binits{S.L.}},
\bauthor{\bsnm{{Palenzuela}}, \binits{C.}}:
\batitle{{Black hole dynamics in Einstein-Maxwell-dilaton theory}}.
\bjtitle{\prd}
\bvolume{97}(\bissue{6}),
\bfpage{064032}
(\byear{2018})
\doiurl{10.1103/PhysRevD.97.064032}
{\href{https://arxiv.org/abs/1706.09875}{{arXiv:1706.09875}}}
{[gr-qc]}
\end{barticle}
\endbibitem

\bibitem[\protect\citeauthoryear{Stein and Joyner}{2005}]{stein:2005ab}
\begin{botherref}
\oauthor{\bsnm{Stein}, \binits{W.}},
\oauthor{\bsnm{Joyner}, \binits{D.}}:
{SAGE}: System for algebra and geometry experimentation
\textbf{39}(2),
61--64
(2005)
\end{botherref}
\endbibitem

\bibitem[\protect\citeauthoryear{{Gerosa} et~al.}{2016}]{gerosa:2016cq}
\begin{barticle}
\bauthor{\bsnm{{Gerosa}}, \binits{D.}},
\bauthor{\bsnm{{Sperhake}}, \binits{U.}},
\bauthor{\bsnm{{Ott}}, \binits{C.D.}}:
\batitle{{Numerical simulations of stellar collapse in scalar-tensor theories
  of gravity}}.
\bjtitle{Classical and Quantum Gravity}
\bvolume{33}(\bissue{13}),
\bfpage{135002}
(\byear{2016})
\doiurl{10.1088/0264-9381/33/13/135002}
{\href{https://arxiv.org/abs/1602.06952}{{arXiv:1602.06952}}}
{[gr-qc]}
\end{barticle}
\endbibitem

\bibitem[\protect\citeauthoryear{{Minazzoli} and
  {Wavasseur}}{2024}]{minazzoli:2024ar}
\begin{botherref}
\oauthor{\bsnm{{Minazzoli}}, \binits{O.}},
\oauthor{\bsnm{{Wavasseur}}, \binits{M.}}:
{Schwarzschild black-hole immersed in uniform electric or magnetic backgrounds
  in Entangled Relativity}.
arXiv e-prints,
2407--17846
(2024)
\doiurl{10.48550/arXiv.2407.17846}
{\href{https://arxiv.org/abs/2407.17846}{{arXiv:2407.17846}}}
{[gr-qc]}
\end{botherref}
\endbibitem

\end{thebibliography}

\end{document}